\RequirePackage{fix-cm}
\documentclass[twocolumn]{svjour3}          

\smartqed  
\usepackage{amsmath}
\usepackage{amssymb}
\usepackage{graphicx}

\begin{document}

\title{A model for inertial particles in curvilinear flows}

\author{Mike Garcia \and Sumita Pennathur}

\institute{M. Garcia \at \email{mikegarcia@ucsb.edu} \and S. Pennathur \at \email{sumita@ucsb.edu}}

\date{Received: date / Accepted: date}

\twocolumn[
  \begin{@twocolumnfalse}
 
\maketitle

\begin{abstract}
The recent advent of advanced microfabrication capabilities of microfluidic devices has driven attention towards the behavior of particles in inertial flows within microchannels for applications related to the separation and concentration of bio-particles. The phenomena of inertial focusing has been demonstrated to be a robust technique in such applications, where the flow of particles in a curvilinear geometry has proven to be particularly advantageous, not only because the geometry can reduce the foot-print of a lab-on-chip device, but also because the coupling of secondary Dean flows to inertial forces allows for exquisite particle manipulations. However, the ability to design a curvilinear channel for a specific application is often based on empirical results, as theoretical models to date typically do not include the effects of a finite sized particle within the flow. Here we present a complete numerical model that directly simulates a particle within a confide curvilinear flow and using this model we investigate the three dimensional focusing behavior of inertial particles as well as the applicability of the point particle assumptions previous researchers have proposed. Finally, we propose a new model that takes into account the full physics, but relies on a perturbation expansion of the lateral forces, where the perturbation parameter is the curvature ratio of the channel. This simple model can be used to predict the behavior of particles in complex channel geometries where the curvature may not be constant.   

\end{abstract}

  \end{@twocolumnfalse}
]


\section*{Introduction}
\label{intro}
It has long been known that particles flowing at a finite Reynolds number ($Re$) can passively migrate laterally across streamlines and focus at stable equilibrium locations \cite{Segre1961} within confined systems as a result of  nonlinear fluid stresses on the particle \cite{Ho1974}. Recently, this phenomena has received a new found interest due to its use in the precise manipulation of micron-sized particles in a continuous microflow, coining the term inertial microfliudics. On the basis of inertial microfluidics, researchers  designed many unique devices to isolate \cite{Nathamgari2015}, sort \cite{Sarkar2016}, focus \cite{Gossett2009,Wang2017} and concentrate \cite{Martel2015} particles. By far the most common device designs leverage curvilinear channels to produce a transverse Dean flow, not only affording a compact design but also allowing for exquisite control of particle streams by simply tuning the Dean forces.  Dean forces arise from the curvilinear geometry which introduces a centrifugal acceleration component directed radially outward as flow navigates through the curved channel. The resulting Dean flow is orthogonal to the streamwise flow direction and is composed of two symmetric counter rotating vortices known as Dean vortices (Fig. \ref{fig:5-1}a). The effect of these vortices in combination with inertial forces serve to perturb the inertial equilibrium locations of a particle into a size dependent stream thus allowing for sorting, concentrating and/or isolating certain kinds of particles. The magnitude of this perturbation is set by the strength of these vortices, which is dictated by the Dean number ($De$) \cite{Nivedita2017,Norouzi2013,Dean1927}.

Inertial Dean flow focusing has been used with both alternating curves and spirals for various bio-analytic purposes \cite{Wang2017,Dicarlo2007,Martel2013,Ozbey2016,Lee2013,Bhagat2008}. However, modeling the flow in these devices for a specific application is quite challenging as the full Navier-Stokes equations are needed to solve for the particle dynamics in these complex channels. Often complete models are too computationally burdensome to be of any practical use in designing these devices \cite{Pedrol2018}. Given the complexity of simulating particle migration, some authors have proposed the use of lattice Boltzmann methods (LBM) as the technique very computationally efficient \cite{Chun2006,Yuan2018}. However, LBM is prone to instability issues because of the coarse grained representation of the fluid-boundary interface \cite{Yuan2018}. By far the most common approach has been a point particle model, where the inertial forces are solved for in a straight channel and the Dean flow effects are added independently by assuming that is is simply a Stokes drag associated with an underlying Dean velocity. This model has been used widely in recent studies, but has been generally limited to small particles and slow flows \cite{Martel2013,Ozbey2016,Zhang2014,Rasooli2018,Martel2013b}. While this approach is quick and has shown some success,  the ability to superpose these two forces may not hold under extreme flow regimes. In particular, this model becomes questionable at high $De$ where the Reynolds number based upon the average Dean flow velocity ($Re_D$) approaches unity (Fig. \ref{fig:5-1}b) and inertial corrections to Stokes drag are necessary. Furthermore, at higher $De$, there is also a redistribution of the axial flow profile (Fig.~\ref{fig:5-1}c) that can alter the shear gradient lift forces. Recently, Dinler and coworkers \cite{Dinler2018} have proposed the use of a direct numerical simulation (DNS) model, where the flow problem is solved in reference frame fixed to a moving sphere similar to \cite{Dicarlo2009,Dinler2018,Martel2013b,Kim2016}. This method is robust and provides the inertial force distribution over the particle in a section of the channel. This method is well suited for fundamental studies \cite{Dicarlo2009}, but not for practical design because the it is computationally inefficient. It is no surprise then that Diner \textit{et al.} applied this model in a curvilinear geometry using coarse parameters and an incomplete description of the momentum equations \cite{Dinler2018}.
\\
\indent There is a need for a simple and precise model that can reliably predict the behavior of confined inertial particles across a wide range of flow parameters in a curvilinear geometry. To address this need we first use a numerical model similar to Dinler \textit{et al.}\cite{Dinler2018}, but here we include Coriolis and centripetal terms in our momentum equations. Based on our numerical observations, we then develop a perturbation based model to predict the lateral forces acting on a spherical particle migrating in a curved channel. We then validate this model against previously published experiments, and compare to the Stokes drag model proposed in the past where for the first time we explicitly demonstrate the break down of the Stokes model. Finally, we use the perturbation based model to design a spiral channel and speculate on how this model can be used to design devices in the future.


\section*{Numerical Model}
\begin{figure}[b]
\centerline{\includegraphics[width=8.5cm]{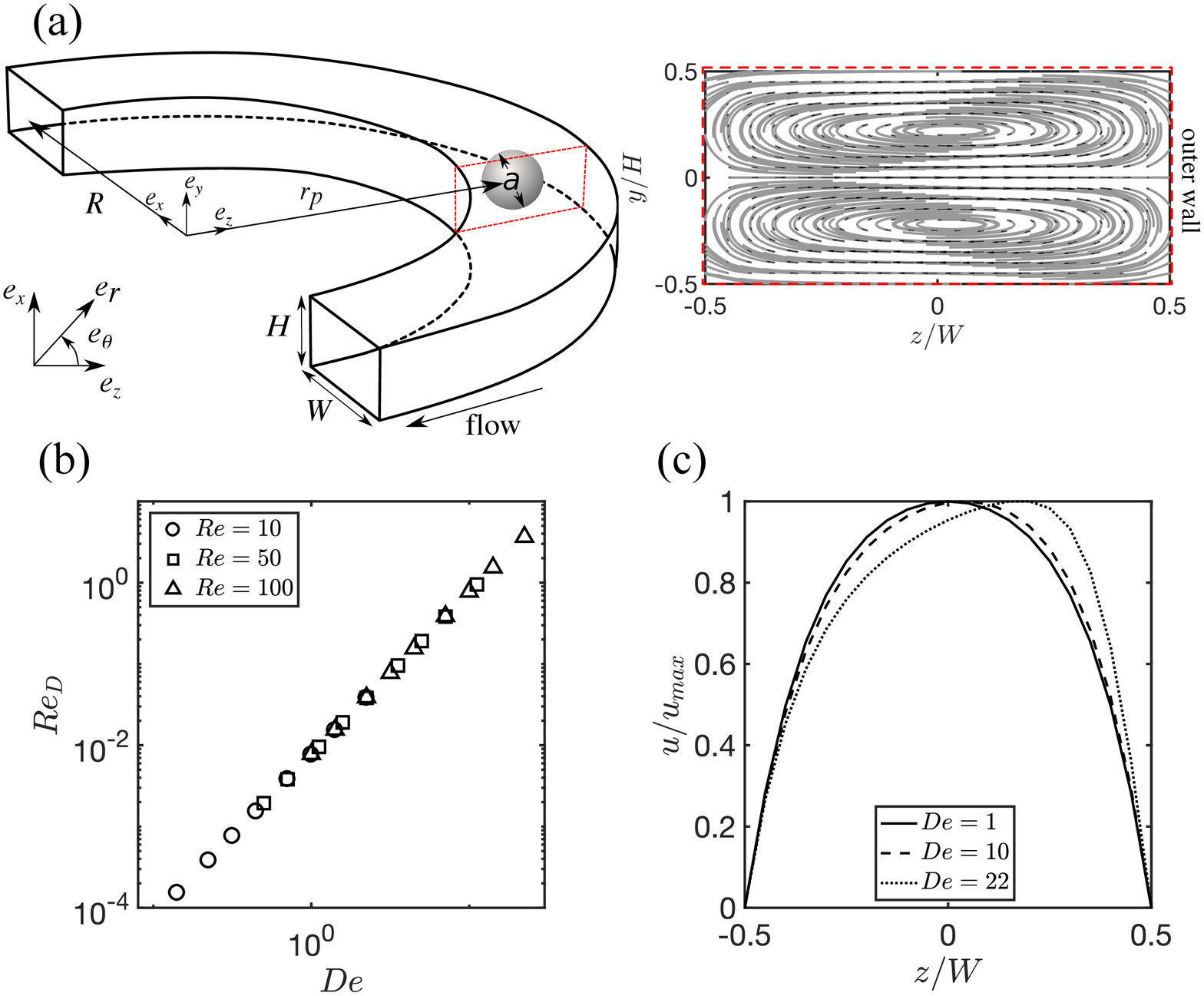}}
\caption{\label{fig:5-1}(a) Schematic illustration of the channel considered in this report. The channel is rectangular with cross-section ($W \times H$) and average radius $R$. The spherical particle of diameter $a$ flows within the confines of the bounding walls at a location $\textbf{r}_p$ relative to origin. A cross sectional slice of the channel reveals that the recirculating flow patterns shown in the red dashed window. (b) A plot of the Reynolds number ($Re_D$) of this recirculating flow versus the Dean number ($De$). For high $De$ the flow has appreciable inertia as the $Re_D$ is $\mathcal{O}(1)$.(c) A plot of the axial flow profile for various $De$. For low $De$ we observe a symmetric profile similar to flow in a straight channel, but for high $De$ the symmetry vanishes due to increased flow redistribution associated with the Dean flow.}
\end{figure}

In order to understand how particles focus in a curved channel at moderate Re numbers we first define a model system. Our model focuses on the flow of a neutrally buoyant particle of diameter $a$ in a channel of rectangular cross-section $W \times H$ ($W/H = 2$), arc-length $5W$ and average radius $R$ (Fig.~\ref{fig:5-1}a). The particle is translating with at a velocity $\textbf{U}_P= -U_p \textbf{e}_\theta = U_p [-\cos\theta \mathbf{e}_x, 0 \mathbf{e}_y, \sin \theta \mathbf{e}_z]  $ and is rotating at an angular velocity $\boldsymbol{\Omega}$ in a flow of average velocity $U$. We define the channel Reynolds number as $Re = \rho U D_h / \mu$  the relative curvature of the channel as $\delta = D_h/2R$, and the Dean number as $De=Re\sqrt{\delta}$, where $\rho$ and $\mu$ are the fluid density and viscosity respectively and $D_h = 2(W+H)/(WH)$ is the hydraulic diameter of the channel. 

To solve for the flow field and pressure around the particle, it is convenient to consider a rotating frame of reference such that the particle appears stationary. The rotating reference frame is a non-inertial frame of reference and thus the Navier-Stokes equations must adopt a form that takes into account the effects of both centripetal and Coriolis forces. Note that we assume a quasi-steady model to eliminate time dependace from the equations:
\begin{equation} \label{eq5-1}
\rho\bigg( \mathbf{u} \cdot \nabla \mathbf{u} + 2 \dot{\boldsymbol{\theta}} \times \mathbf{u} + \dot{\boldsymbol{\theta}} \times \dot{\boldsymbol{\theta}} \times \mathbf{r}\bigg)  = \mu \nabla^2 \mathbf{u} - \nabla p\\
\end{equation}
\begin{equation} \label{eq5-2}
   \nabla \cdot  \textbf{u} = 0
\end{equation}
where $p$ is the fluid pressure field, $\textbf{u}$ is the fluid velocity field in the rotating reference frame, $\dot{\boldsymbol{\theta}}$ is the angular velocity of the frame, and \textbf{r} is the position vector of a fluid element about the point of rotation of the frame. The frame velocity, $\dot{\boldsymbol{\theta}}$ is related to the particle velocity by:  $\textbf{U}_p = \textbf{r}_p \times \dot{\boldsymbol{\theta}}$, where $\textbf{r}_p$ is the position vector of the particle center relative to the point of rotation (\textit{i.e.} at the origin).

The translational and rotational flow rates of the suspended particle ($U_P$ and $\boldsymbol{\Omega}$, respectively) can be self-consistently determined by setting conditions such that the axial motion satisfies a drag constraint $F_{\theta} = 0$ and its rotational motion satisfies a torque constraint $\tau_{r} = \tau_{z} = \tau_{\theta} = 0$. The boundary conditions of this problem are in the rotating reference frame, therefore, the no slip condition on the walls is, $\textbf{u}_{wall}  = -\dot{\boldsymbol{\theta}} \times \textbf{r}$.
The no slip condition on the particle is enforced by assigning a velocity to the surface of the sphere corresponding to that of a rigid body rotation at angular velocity, $\textbf{u}_{surface} = \boldsymbol{\Omega} \times (\textbf{r}-\textbf{r}_{p})$.

Far from the particle the flow is undisturbed and regains the behavior of flow in the absence of a particle. To solve for the unknowns (\textit{i.e.}, $\mathbf{u}$, $p$, $U_{p}$ and $\boldsymbol{\Omega}$) we couple the Navier-Stokes equations to the equations constraining the particle motion (\textit{i.e.} torque and force free equations of motion) and numerical solve directly using the COMSOL multiphysics software. This procedure is performed for a lattice of discrete positions of the particle within the symmetric top half of the cross-section of the channel. To calculate the lift force on the particle, we integrate the surface stresses on the particle in the appropriate direction ($y$ or $z$). Note that because $a/R << 1$ and the particle is simulated at $\theta = 0$, we can say that $\textbf{e}_r \approx \textbf{e}_{z}$ and $\textbf{e}_{\theta} \approx \textbf{e}_{x}$ for the purposes of integrating the hydrodynamic stresses on the the surface of the particle. The numerical model presented in this report investigates the steady state forces $\textbf{F}_{DNS}$ on a finite sized particle through direct numerical simulation of the flow field.
\begin{equation}  \label{eq5-12}
\textbf{F}_{DNS} = \int_s  \textbf{n} \cdot \textbf{T} \,ds - m_p \dot{\boldsymbol{\theta}} \times (\dot{\boldsymbol{\theta}} \times \textbf{r}_p)
\end{equation}
Here the first term on the right hand side of the equation represents  the hydrodynamic forces, where $\textbf{T} = \mu \nabla^2\textbf{u} - p\textbf{I}$, is the total stress tensor of the flow around a particle that is restricted from moving laterally. The second term represents the contribution of the centripetal acceleration on the particle. This numerical model includes finite size effects of the particle, the redistribution of the axial velocity profile and the Coriolis and centripetal acceleration terms in the momentum equation. 

Computational modeling efforts were performed using COMSOL multiphysics software (version 5.2a) using a 3D CFD model using a model with $6 \times 10^5$ degrees of freedom. To calculate inertial lift forces we coupled the equations of fluid motion to a set of global differential equations to solve for the translational and rotational velocity of the particle. The Coriolis and centripetal terms in the Navier-Stokes equations were modeled as a body force. The drag on the particle ($F_{\theta} \approx F_x$) was calculated in COMSOL by integrating the total stress over the surface of particle in the axial direction ($\textbf{e}_\theta \approx \textbf{e}_x$). Similarly the torque on the particle ($\boldsymbol{\tau}$) was calculated by integrating the differential torque ($d\boldsymbol{\tau} = (\textbf{r}-\textbf{r}_p) \times \textbf{n} \cdot \textbf{T} ds$) on the surface of the particle. A mesh sensitivity analysis was conducted to show that calculated lift forces were independent of mesh density to within 1\% error.


\section*{Numerical Results}
\begin{figure}
\centerline{\includegraphics[width=8cm]{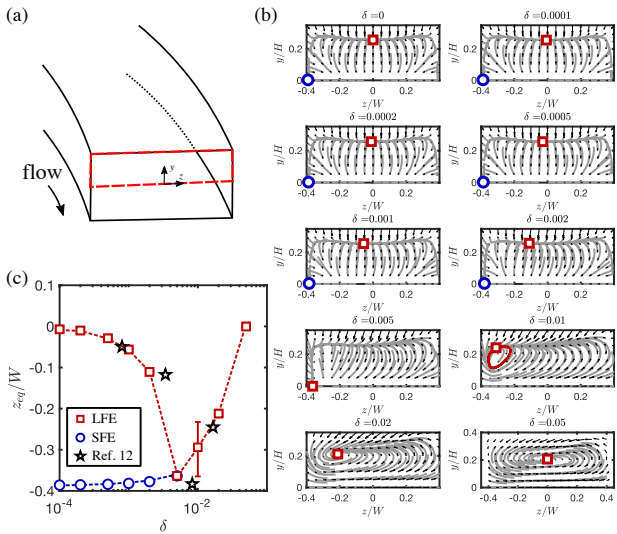}}
\caption{\label{fig:5-2}(a) Schematic illustration of a curved channel depicting the region of interest (red dashed box).(b) The cross-section plots show the simulated resultant force $\textbf{F}_{DNS}$ on the particle for multiple channel geometries $(\delta = D_H/2R)$ at $Re = 100$. The gray line are streamlines of the force field and are for visualization purposes. The red square denotes the location of the long face equilibrium (LFE), the blue circle and green diamond denote the short face equilibrium (SFE). Note that only the top half of the channel is shown due to symmetry. (c) Stable equilibrium location for both LFE and SFE as a function of the relative channel curvature $\delta$ for the results of this numerical model and experiments done by Martel \textit{et al.}, 2013 \cite{Martel2013}; only the inner SFE is shown for clarity}
\end{figure}
Fig. \ref{fig:5-2}a shows a schematic illustration of the top half of the channel cross-section over which we simulate a particle spanning the parameters $Re = 10$ to $100$ and $\delta = 0$ to $0.05$. Fig. \ref{fig:5-2}b shows the force-field $\textbf{F}_{DNS}$ for a subset of the simulation space, specifically an intermediate sized particle ($a/D_h = 0.150$) and at $Re = 100$. Under these conditions, and without loss of generality, we observe that the force-fields are progressively perturbed for increasing channel curvature (\textit{i.e.} $\delta$) at a constant flow rate ($Re=100$) (Fig. \ref{fig:5-2}b). Further, in a straight channel (\textit{i.e.} $\delta = 0$) we see four stable equilibrium locations, where the equilibrium along the long faces (LFE) attracts more streamlines than the equilibrium along the short faces (SFE). The phenomena of a relatively more stable LFE has been observed experimentally and numerically for a rectangular channel under the similar conditions \cite{Liu2015}. As the channel curvature increases, the location of the LFE (red square) shifts towards the inner wall. The LFE eventually merges with the inner SFE (blue circle) at sufficiently high channel curvature ($\delta = 0.005$). After this point the SFE/LFE begins a retrograde motion towards the outer wall (Fig. \ref{fig:5-2}c). Interestingly, after the SFE/LFE switch direction, the equilibrium destabilizes. At this point the particle is not focused at a single point, but rather orbits in plane (Fig. \ref{fig:5-2}b, $\delta = 0.01$). These results are compared to the experimentally obtained values of the LFE and show excellent agreement \cite{Martel2013}.  Note that there is also a SFE that corresponds to the outer wall, however it is not a stable equilibrium location after $\delta = 0.005$ and has been neglected for the clarity of this discussion.

The non-monotonic shift in LFE at a fixed $Re$ for varying $\delta$ is caused by the presence of the Dean flow within the channel \cite{Martel2013}. Initially, for low $\delta$ the LFE is at a vertical location where locally the Dean flow is directed towards the inner wall. The strength of this Dean flow increases with the curvature of the channel (Fig. \ref{fig:5-1}b) and thus the LFE shifts towards the inner wall with increasing $\delta$. As LFE the shifts towards the inner wall the Dean flow in that region beings to impart a vertical force that is directed in the negative $y$-direction (Fig. \ref{fig:5-1}a). This causes the LFE to move towards the inner SFE and eventually merge. Finally, the merged LFE/SFE migrate towards the outer wall (locally the direction of the Dean flow) at sufficiently high $De$. This transition occurs because the shear gradient across the width of the channel on the inner half of the channel is insufficient to counter the increasing Dean flow forces; thereby adjusting the location of the LFE/SFE towards the center-line \cite{Martel2013}. 

\section*{Second Order Model (SOM)}

\begin{figure*}[h]
\centerline{\includegraphics[width=15cm]{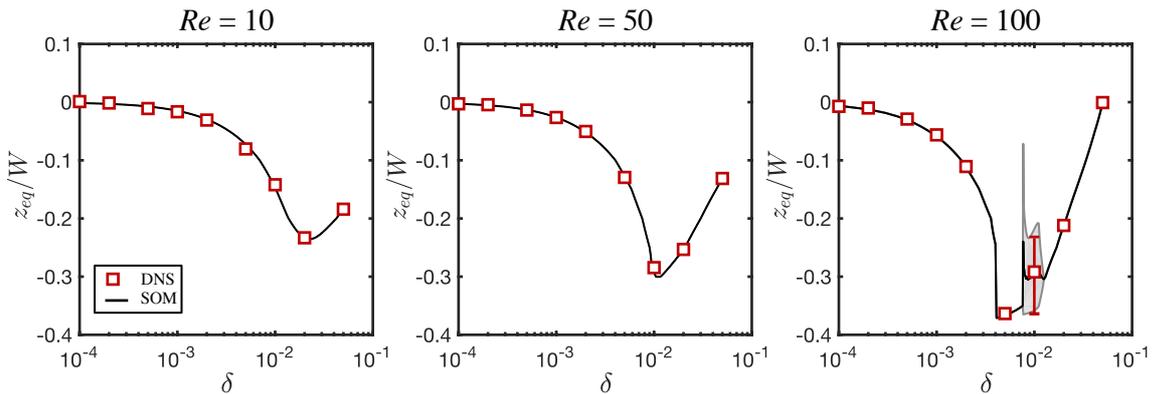}}
\caption{\label{fig:5-3} Stable equilibrium location as a function of the relative channel curvature $\delta$ for three distinct $Re$ and $a/D_h = 0.150$. The square markers represent results from direct numerical simulations (DNS) and the solid lines represent the results from the second order perturbation model (SOM). The shaded region at $Re=100$ represents the orbit focusing limits.}
\end{figure*}

 The process of solving for the inertial forces with the numerical model proposed in the previous subsection is computationally intensive and thus difficult to apply as an optimization and/or design tool. Therefore, we developed a second order model that can produce quantitative results with significantly less computational power such that we can use the model to design systems for particular applications. This model follows the work of Dean \cite{Dean1927}, and is based on the observation that the inertial forces are increasingly perturbed for increasing channel curvature (Fig. \ref{fig:5-2}). Dean's seminal study laid the framework to describe flow in a curved pipe with pertubation method based analytic solutions with the curvature ratio as the perturbation parameter. Following this work, we propose a similar model, which assumes instead that the forces on a particle (and not the flow) in a curved geometry can be thought of as a perturbation series. Like Dean's model the leading term in this power series is the solution of the straight channel problem, while further terms describe the deviation in the solution due to increased curvature $\delta$. 
 
 We first consider a perturbation of the lateral lift forces $\textbf{F}_{DNS}$ about the $\delta = 0$, \textit{i.e.} straight channel case. 
\begin{equation} \label{eq5-14}
    \textbf{F}_{DNS} = \textbf{F}_0 + \delta \textbf{F}_1 + \delta^2 \textbf{F}_2 + {\mathcal{O}}(\delta^3)
\end{equation}
Where $\textbf{F}_{0} \equiv \textbf{F}_{DNS} \big|_{\delta = 0}$ is the full physics lift force calculated for a particle under a given $Re$ for a straight channel (\textit{i.e.}, $\delta = 0$), $\textbf{F}_{1}$ and $\textbf{F}_{2}$ represents the effects of channel curvature on the lateral forces experienced by a particle. We speculate that for sufficiently small $\delta$, $\textbf{F}_{1}$ and  $\textbf{F}_{2}$ in Eq. \ref{eq5-14} are the only terms required to model the lateral forces and thus we can neglect any higher order terms. Here the objective is to obtain quantitatively precise forces values with minimal computational requirement and thus we truncate the infinite series after only three terms. In this work, we do not try to analytically identify the form of the functions $\textbf{F}_{1}$ and  $\textbf{F}_{2}$, but explore how it can be constructed by using a minimal set of full physics simulations. We show below that $\textbf{F}_{1}$ and  $\textbf{F}_{2}$ (and hence, $\textbf{F}_{DNS}$) can be reliably constructed using just three full physics simulations - to do so we solve for these perturbation functions by rewriting Eq.~\ref{eq5-14} for a fixed $Re$ and $a/D_h$. This approach has been previously demonstrated in our previous work and has shown excellent agreement with both experimental and numerical results \cite{Garcia2018}.

\begin{equation} \label{eq5-17}
 \textbf{F}_{1} = \frac{\big(\delta_1^2\textbf{F}_0 - \delta_2^2\textbf{F}_0 - \delta_1^2\textbf{F}_{DNS}\big|_{\delta = \delta_2}+\delta_2^2\textbf{F}_{DNS}\big|_{\delta = \delta_1}\big)}{\delta_1(\delta_2^2-\delta_1\delta_2)}
 \end{equation}
 \begin{equation} \label{eq5-18}
 \textbf{F}_{2} = -\frac{\big(\delta_1\textbf{F}_0 - \delta_2\textbf{F}_0 -\delta_1\textbf{F}_{DNS}\big|_{\delta = \delta_2} + \delta_2\textbf{F}_{DNS}\big|_{\delta = \delta_1}\big)}{\delta_1(\delta_2^2-\delta_1\delta_2)} 
 \end{equation}
 Here $\textbf{F}_{DNS}\big|_{\delta = \delta_1}$  and $\textbf{F}_{DNS}\big|_{\delta = \delta_2}$ are the full physics simulation results for flow at the same $Re$ in two distinct channels of curvature ratio $\delta = \delta_1$ and $\delta = \delta_2$ respectively.

To demonstrate the utility of such a model, we calculate $\textbf{F}_1$ and $\textbf{F}_2$ using Eq. \ref{eq5-17} and Eq. \ref{eq5-18} with only three DNS ($\delta = 0, 0.02, 0.05$) at a fixed $Re$ and $a/D_h$. Fig. \ref{fig:5-3} shows the results of this model, where we show the predicted equilibrium location as a function of $\delta$ and compare to the discrete DNS results for three distinct flow regimes ($Re$). Here we use the equilibrium location as a concise representation of the more complex force maps. From this figure, it is apparent that the SOM reconstructs the lateral lift forces well with little discernible error; with the advantage of the SOM being that it only requires knowledge of three full simulations as opposed to the nine DNS shown in the figure. Moreover, the model is not limited to discrete values of $\delta$ - as it can predict the particle behavior at any combination of $\delta$ or $Re$ provided that basis are known. The second order model is so precise, in fact, that it even predicts the orbit focusing for $Re =100, \delta =0.01$ that was observed previously in Fig. \ref{fig:5-2}b; It does so with no knowledge of the flow as $\delta = 0.02$ and $\delta = 0.05$ were used to solve for the model parameters (Fig. \ref{fig:5-3}, $Re=100$).  

\section*{Discussion}
\begin{figure*}
\centerline{\includegraphics[width=15cm]{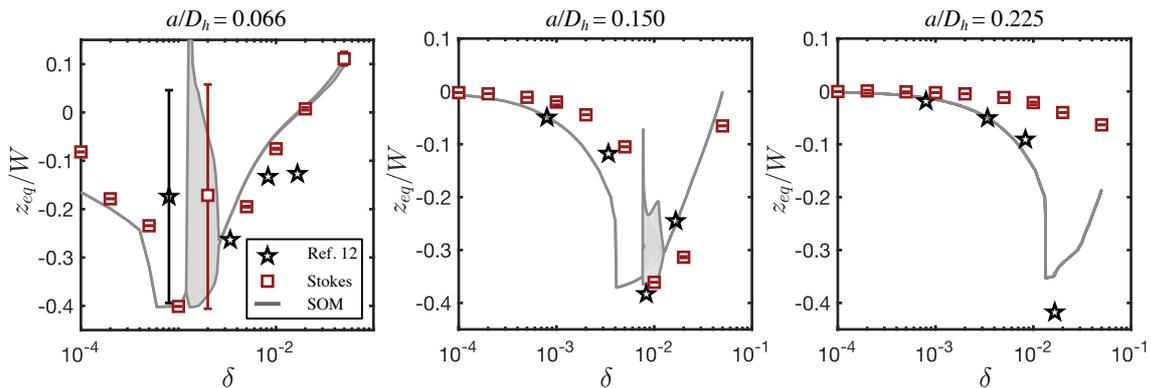}}
\caption{\label{fig:5-4} Stable equilibrium location as a function of the relative channel curvature $\delta$ for three distinct particle sizes ($a/D_h$) at $Re = 100$. The square markers indicate the predictions from a simple Stokes drag model (Stokes). The solid lines are the predictions from the second order model (SOM). The stars represent the experimental results from Martel \textit{et al}.  \cite{Martel2013}.}
\end{figure*}

As mentioned in the introduction, the Stokes model has been proposed in previous studies as a quick an reliable method for modeling the lateral forces on a particle in curvilinear channel. However, the effectivneness of such a model has yet to be demonstrated particularly across a wide range of particle sizes. Here we compare the results of our SOM with the simple Stokes model and experimental results of Martel \textit{et al.} \cite{Martel2013} to determine under what conditions either model is valid. As a reminder, the Stokes model adds the inertial lift forces ($\textbf{F}_0$), derived for a straight channel, with a force caused by the local Dean flow velocity ($\textbf{U}_{Dean}$) in the channel. Here $\textbf{U}_{Dean}$ is the lateral flow field in a curved channel with no particle at discrete values of $\delta$ and $Re$. It is important to note that this approach is also computationally inefficient as it requires knowledge of the underlying flow field, which in general can be spatially varying.

 \begin{equation} \label{eq5-19}
\textbf{F}_{Stokes}= \textbf{F}_0 + 3\pi \mu a \textbf{U}_{Dean}
 \end{equation}
Here the centripetal force term has not been included in this Stokes model and in previous work \cite{Martel2013,Ozbey2016,Zhang2014,Rasooli2018,Martel2013b}. It is a serendipitous occurrence and can be shown that for a small and neutrally buoyant particle that the pressure gradient term associated with the undisturbed flow imparts a force that exactly cancels out centripetal forces \cite{Maxey1998,Lim2003}. While this Stokes model has been proposed as a simple tool and used heavily in literature, It is not obvious that it should provide meaningful results for flows with large particles and at high $Re$.

Fig. \ref{fig:5-4} shows a comparison of the predicted focusing location for the two models discussed in this article with the experimental results of Martel \textit{et al.} \cite{Martel2013}. The SOM agrees well with all experimental results. Using the SOM we can precisely replicate the experimental results to see that in general a small addition of curvature causes the focusing location of a particle to shift towards the inner wall. Interestingly, for the smallest particles ($a/D_h = 0.066$ and $a/D_h = 0.150$), at a sufficiently high curvature, we observe that the particles can be entrained in an orbit rather than having a single focusing location, a result that is confirmed by experiments by Martel \textit{et al.} \cite{Marte2013}. As expected, for small particles the Stokes model and SOM agree well, but for larger particles and at higher $\delta$, the predicted focusing locations begin to diverge. This discrepancy can be attributed to two factors: 1) the redistribution of the axial flow profile at high Dean number (Fig. \ref{fig:5-1}a, $De= Re\sqrt\delta$) and 2) finite size effects which are not considered by the point particle assumption inherent in the Stokes model. Our findings resolve confusion about the size dependence of inertial lift forces combined with Dean flow experienced by particles traveling through curved microchannels. Many studies have assumed that this behavior can be represented by a simple Stokes model. However, by numerically dissecting the equations of fluid flow around the particle, we find that this assumption does not hold for larger particles. This result is of particular significance in many biological applications when the particles of interests are cells which often are large compared to the size of the confining channel.   

Finally, to demonstrate one potential application of the SOM, we consider the focusing of particles in a ``spiral channel''. The spiral channel is a geometry that is ubiquitous in inertial microfluidics. This geometry has been utilized in numerous studies to manipulate particles \cite{Nivedita2017,Lee2013,Bhagat2008,Martel2013,Martel2012}. However, modeling the focusing behavior of particles in this type of channel is typically quite challenging. The challenge is due to the fact that the channel does not have a single radius of curvature, but rather a radius of curvature that is evolving with the streamwise direction. Modeling the the trajectories of particles in this type of channel using the techniques outlined in the introduction of this article would be not be practical. The full 3D geometry has a very large aspect ratio and thus the computational time and memory requirements would be extensive. However, the SOM is well suited for this problem because it predicts the local force values using only $\delta$ as the input parameter (for a given flow). Thus providing precise force predictions with no knowledge of the flow field everywhere in channel or long computational efforts.

\begin{figure}
\centerline{\includegraphics[width=8cm]{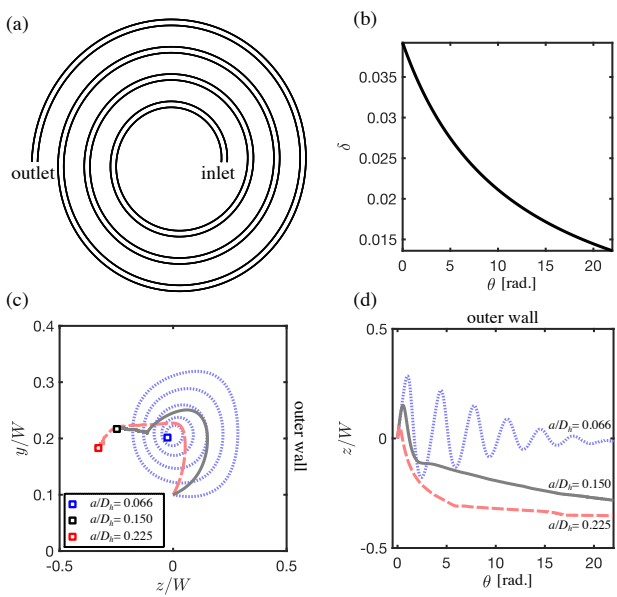}}
\caption{\label{fig:5-5} (a) Schematic illustration of the spiral channel considered in this chapter. (b) The radius of curvature in this spiral channel decreases in the stream-wise direction as $\delta \sim 1/\theta$. at the inlet $\delta = 0.392$ and at the outlet $\delta = 0.136$. (c) The cross-sectional trajectories of the three particles in this spiral channel at $Re=100$. The particles are seeded at a common reference and their outlet location is indicated by the square markers. (d) A projection of the particle trajectories in (c) onto the stream-wise plane from inlet ($\theta = 0$ to outlet $\theta = 7 \pi$).}
\end{figure}

Here we consider an Archimedean spiral (Fig. \ref{fig:5-5}a) with a similar cross-section as the previous section (\textit{i.e.} $W/H = 2$) that has a radius of curvature that is parameterized by $R = a + b\theta$. Where $R$ is the local channel radius, $a$ is the channel radius at the inlet ($\theta = 0$) and $b$ is a parameter that controls the spacing successive between spirals. To determine the lateral forces on a given particle, we first use the expression for $R$ to to derive and expression for relative channel curvature everywhere in the channel:
 \begin{equation} \label{eq5-21}
\delta = \frac{D_h}{2(a + b\theta)}
 \end{equation}
From Eq. \ref{eq5-21} it apparent that curvature can vary significantly over the length of the channel. In Fig. \ref{fig:5-5}b we show this variation in a polar coordinate representation from the inlet to the outlet of this spiral channel. Using this knowledge, we can then compute the lateral forces anywhere in the channel using Eq.\ref{eq5-21} and Eq. \ref{eq5-21}. The trajectories of a given particle are then calculated using a first order time stepping approximation:
\begin{equation} 
 \theta_{n+1} = \theta_{n} +  \frac{u_{\theta}(y_{n}, z_{n})}{R}\Delta t
 \end{equation}
 \begin{equation} 
 y_{n+1} = y_{n} +  \frac{F_{y}(y_{n},z_n)}{3\pi \mu a} \Delta t
 \end{equation}
 \begin{equation} 
 z_{n+1} = z_{n} +  \frac{F_{z}(y_{n},z_n)}{3\pi \mu a} \Delta t 
\end{equation}

Where $u_\theta$ is the streamwise flow field, $F_y$ and $F_z$ are the predicted forces in the lateral directions calculated using the SOM.
Fig. \ref{fig:5-5}c and \ref{fig:5-5}d show the the trajectories calculated for three distinct particles $a/D_h = 0.066,0.150, 0.225$ under the same flow conditions ($Re=100$). In Fig.\ref{fig:5-5}c and \ref{fig:5-5}d we seed three particles at a common location as a basis for comparison ($z/W = 0, y/H = 0.1$). Interestingly, we see that the particles never reach an equilibrium, but rather are constantly migrating (Fig. \ref{fig:5-5}d). This result is rationalized by considering that the curvature is never constant and thus the forces on the particles are perpetually evolving. These results agree well with experimental findings, where the focused particle streaks in a similar spiral channel were seen to continuously migrate \cite{Martel2012}. Furthermore, we note that the trajectory is high oscillatory for smaller particles ($a/D_h = 0.066$), but the oscillations dampen towards the outlet. Suggesting that smaller particles in this particular geometry may take a considerable channel length to actually focus. Another intriguing observation of this specific channel is that under this configuration we actually observe quite significant separation of the focused particle streams, suggesting that this may be a viable channel for separation purposes.  
It is clear that there is tremendous value in predicting the lateral forces in an arbitrary geometries such as the spiral channel presented here. One could imagine easily iterating over thousands of channels to obtain the optimal design for separating particles in minutes. The SOM presented here is not limited to spiral channels, but can easily be adapted to any channel where the local channel curvature can be parameterized such as in a serpentine channel.  Furthermore, our SOM can be used to better understand the complex focusing dynamics observed in many previous studies \cite{Martel2013,Martel2012}. 

\section*{Conclusion}
There is a clear need for a simple yet precise model of the forces behind the motion of particles in moderate Reynolds number flows within curved channels. This is a first attempt to precisely model the equations of fluid motion to determine the effect of channel curvature on the behavior of inertial particles. Using the full numerical model we observed that particle equilibrium locations are highly dependent on the magnitude of the underlying Dean flow. Based on this full model we have developed a second order model that provides a simple yet precise representation of these forces with minimal computation burden. We have demonstrated that this second order model is both more precise and versatile than the commonly referenced Stokes model. Future work in this problem will answer the ill-posed inverse problem for which there is no tractable solution \textit{i.e.} can a channel be designed given a desired particle focusing location? Continued development and investigation of this model can help answer this question and make  these  results  more accessible to researchers with no knowledge of inertial microfluidics.

\section*{Acknowledgements}
The authors would like to thank Professor Paolo Luzzato-Fegiz for insightful discussion regarding rotating reference frames. M.G. was funded by XXX.

\end{document}